\documentclass[runningheads]{llncs}
\usepackage[T1]{fontenc}
\usepackage{graphicx}
\usepackage{relsize} %To get big sigmas in equations
\usepackage{orcidlink} % To add ORCID ids as little images
\usepackage{url}
\usepackage{amsmath}
\usepackage{amsfonts} 
\DeclareMathOperator{\LeakyReLU}{LeakyReLU} %for maths formulas (non italic txt)
\DeclareMathOperator{\ReLU}{ReLU} %for maths formulas (non italic txt)
\DeclareMathOperator{\softmax}{softmax} %for maths formulas (non italic txt)
\usepackage{color}

\usepackage[normalem]{ulem}

\begin{document}
\title{Graph-based multimodal multi-lesion DLBCL treatment response prediction from PET images}%\thanks{Supported by organization x.}}

% %%%% ANONYMISED
\author{Oriane Thiery\inst{1}\orcidlink{0009-0003-1314-1988} \and
Mira Rizkallah\inst{1}\orcidlink{0000-0001-7724-9304} \and
Clément Bailly\inst{2,3}\orcidlink{0000-0001-8313-3287} \and
Caroline Bodet-Milin\inst{2,3}\orcidlink{0000-0002-8219-3592} \and
Emmanuel Itti\inst{4}\orcidlink{0000-0003-1578-4058} \and
René-Olivier Casasnovas\inst{5}\orcidlink{0000-0002-1156-8983} \and
Steven Le Gouill\inst{3}\orcidlink{0000-0001-9840-2128} \and
Thomas Carlier\inst{2,3}\orcidlink{0000-0002-6932-7322} \and
Diana Mateus\inst{1}\orcidlink{0000-0002-2252-8717}}
\authorrunning{O. Thiery et al.}
\titlerunning{Graph-based multimodal multilesion DLBCL response prediction from PET}
\institute{Nantes Université, Centrale Nantes, CNRS, LS2N, UMR 6004, France \and
Nuclear Medicine Department, University Hospital,  Nantes, France \and Nantes Université, Inserm, CNRS, Université d'Angers, CRCI2NA, Nantes, France \and Nuclear Medicine, CHU Henri Mondor, Paris-Est University, Créteil, France \and Hematology, CHU Dijon Bourgogne, Dijon, France}

%Mail
%CB : clement.bailly@chu-nantes.fr
%CBM : caroline.milin@chu-nantes.fr
%EI : emmanuel.itti@aphp.fr
%ROC : olivier.casasnovas@chu-dijon.fr
%SLG : steven.legouill@curie.fr 

\maketitle              % typeset the header of the contribution
\begin{abstract}
%The abstract should briefly summarize the contents of the paper in 150--250 words.

Diffuse Large B-cell Lymphoma (DLBCL) is a lymphatic cancer involving one or more lymph nodes and extranodal sites. Its diagnostic and follow-up rely on Positron Emission Tomography (PET) and Computed Tomography (CT). After diagnosis, the number of non-responding patients to standard front-line therapy remains significant (30-40\%). This work aims to develop a computer-aided approach to identify high-risk patients requiring adapted treatment by efficiently exploiting all the information available for each patient, including both clinical and image data. We propose a method based on recent graph neural networks that combine imaging information from multiple lesions, and a cross-attention module to integrate different data modalities efficiently. The model is trained and evaluated on a private prospective multicentric dataset of 583 patients. Experimental results show that our proposed method outperforms classical supervised methods based on either clinical, imaging or both clinical and imaging data for the 2-year progression-free survival (PFS) classification accuracy.

\keywords{Multimodal data fusion \and Graph Neural Networks \and Cross-attention \and DLBCL \and Treatment Response %\and Survival Analysis 
\and PET.}
\end{abstract}
%
%
%
%\newpage
%
\section{Introduction}%Why+contribution+related works
%pas exactement suivi la proposition de Diana car ça me semblait plus cohérent comme ça, mais je peux faire évoluer ça après des retours

%First give the clinical motivation, define the disease, talk about its survival, prevalence, etc. In a separate paragraph discuss the fact that the diagnosis and follow-up by physicians includes analysing clinical biomarkers and semi-quantitative analysis of full-body FDG18 PET images
%What is the final aim: CAD for the identification of high-risk patients requiring adapted treatment
%•••••••••••••••

%Clinical motivation (disease) :
Diffuse Large B-cell Lymphoma (DLBCL) is a cancer of the lymphatic system and the most common type of Non-Hodgkin Lymphoma (NHL). Its incidence is regularly growing, accounting for 30-40\% of the 77240 new NHL cases %of 
in the US in 2020 \cite{Susanibar-Adaniya_Barta_2021}. %\cite{Liu_Barta_2019,Siegel_Miller_Jemal_2020}. %The disease is usually aggressive and most patients necessitate immediate treatment, with a rapidly growing tumour mass involving one or more lymph nodes and extranodal sites. 
The diagnosis and follow-up 
%of this disease 
include analysing clinical biomarkers and the semi-quantitative interpretation of 18F-Fluorodeoxyglucose (FDG)-PET/CT images. %(\textsuperscript{18}F-Fluorodeoxyglucose positron emission tomography combined with computed tomography) images. %An accurate analysis requires years of training, and nevertheless may be prone to intra- and inter-expert variabilities.
%
%vrai cliniquement ? %Suffisant sur la maladie ?
%Mean survival at five years is around 70\% (je ne retrouve pas la source de cette info)
%J'arrive pas à trouver une survival venant d'une source fiable et récente
%Le taux de survie à 5 ans pour les personnes diagnostiquées entre 2010 et 2015 est de l'ordre de 61 %. %https://www.ellye.fr/lymphome-diffus-a-grandes-cellules-b-lbdgc-ou-dlbcl (suffisant ?)
%Discuss :
 %Some of the most prognostic clinical information include the age of the patient, Ann-Arbor staging, the number of extranodal sites or his concentration of lactate dehydrogenase (LDH).  %Qu'est-ce qu'on peut dire de plus sur ça ? Dire quels types de biomarkers on utilise ? (aaIPI, Ann-Arbor staging, number of extranodal sites, etc (most pronostic ones ?)) % Mentionner des infos sur l'exploitation des images ?
%
%Towards assisting 
To assist such analysis, %most of the 
existing methods in clinical studies focus on clinical data with classical but interpretable methods \cite{gouill2021}. 
%On the other hand, from 
In the image analysis domain, the trend is either to use %full 
deep learning methods %on images 
\cite{Yuan2022MultimodalDL} or to focus on automatically extracting quantitative information (radiomics features) from PET images and combining them with machine learning methods \cite{jiang2022}. 
%
%What is the final aim: CAD for the identification of high-risk patients requiring adapted treatment :
In this context, we aim 
%at developing 
to develop a computer-aided method to identify high-risk patients at diagnosis, relying on both clinical and imaging information. %reformuler selon le sujet de la conf

%Ici on doit noter les differents defis, Lies a la modalité d'image, au fait que la information est concentrée sur multiples lesions qui sont souvent de tres petite taille, etc, Puis finalement, que la plus part de méthodes approchant les tache de la prediction de survie de patients soit se concentrent sur les données cliniques avec de methodes plutot classiques mais interpretables, soit sur les données image :
%Défis:
We face multiple challenges when designing a risk classification approach from heterogeneous multimodal data. %Not only 
First, the quantity of available data on this disease 
%may be considered as 
is often limited.
%, but also 
Also, the information in the PET volumes is spread over multiple typically small lesions, making feature extraction difficult. % lesions whose size is often very small
%Features extracted from multiple sparsely distributed small lesions may not be reliable. 
In addition, both 
%the voxel size 
image resolution and the number of lesions can vary significantly across patients, %which could make the learned models process very hardly generalizable. 
hindering generalizability.
Finally, the integration of the different modalities is still an open question in the field~\cite{Baltrušaitis_Ahuja_Morency_2019}.% \sout{on which no consensus have been found}.%Précise bcp plus les problématiques associées à ce que je soulève

In this paper, we rely on recent advances in Graph Attention Networks (GATs) to combine the information from the multiple lesions while handling the variable number of lesions. We further couple the GAT with a cross-attention fusion module to efficiently integrate data from clinical and imaging modalities. The model is trained and evaluated using a private prospective multicentric dataset with 583 patients suffering from DLBCL. %the multicentric GAINED dataset (NCT 01659099) with 583 patients suffering from DLBCL. 
Experimental validation results show that our proposed method yields a good 2-year progression-free survival (PFS) classification accuracy while outperforming classical supervised methods based on either clinical, imaging or both clinical and imaging data.

\section{Related work}
%In this section, we briefly summarize ..
%\mdf{Mira : Je pense qu'il faut plutot faire une subsection pour single lesion (information manquante) / full image (irrelevant information) et de prendre en compte les différentes lesions (dispersion..). --> l'intêret du graphe (se situer ici). Et dans l'autre sous section parler vers comment incorporer les données cliniques/multimodal fusion et citer les papiers qui font late fusion / early fusion et se situer par rapport à ceux là.}
%\subsection{Automated study of cancer patients' treatment response based on full-body images} %single lesion/corps entier vs graph
Recently, there has been a growing interest in developing computer-assisted methods analysing full-body PET images to support diagnosis and treatment decisions 
%for some cancers' (such as multiple myeloma or DLBCL) 
of oncological patients.  
Different approaches have been considered, relying either on a region of interest (ROI) surrounding a single lesion, or on the full image. 
%For example, survival outcome or prognosis predictions can be obtained from the %region of interest surrounding single lesions lesions ROIs~\cite{ruan2019,Li2019}. 
For example, methods in~\cite{ruan2019,Li2019} make outcome or prognosis predictions from lesions ROIs.
However, images are only part of the patient's %available 
information that physicians rely on to determine the best treatment options. 
%Another approach proposed in \cite{morvan2019leveraging} relies 
Other approaches \cite{morvan2019leveraging} rely on both clinical data and image features from the most intense focal lesion %of patients affected by multiple myeloma for a PFS prediction task. 
to predict the PFS of multiple myeloma patients.
However, for all these methods, resuming a full-body image %and multiple lesions 
to a single ROI may not 
%be completely representative of 
fully represent the patient's state as it overlooks the information from other lesions and their %relative 
potentially structured spatial distribution. 
%\dm{This distribution will be an important }
%\sout{(that we will call the structure of the lesions from this point on)}.

% Blanc-Durand et al. \cite{Blanc-Durand_Campedel_Mule_Jegou_Luciani_Pigneur_Itti_2020} use the entire full-body CT scan and clinical data to estimate progression-free survival (PFS) and overall survival (OS) for patients with non-small-cell lung cancer. %corps entier
% This method is also sub-optimal because it adds noise to the features by considering body regions which are unaffected by the disease.
%  % single lesion

Few papers tackle the problem of incorporating both the imaging descriptors and the underlying structure of all the 
%lesions of a patient 
patient lesions~\cite{Lv_Zhou_Peng_Peng_Lin_Wu_Xu_Lu_2023,Kazmierski_Haibe-Kains_2021,Aswathi_Rizkallah_Frecon_Bailly_Bodet-Milin_Casasnovas_Gouill_Kraeber-Bodéré_Carlier_Mateus_2023}. %They all rely on graph representations to model this structural information and a graph neural network (GNN) on the top of this structure to provide predictions 
They rely on graph representations to model this structural information and build a graph neural network (GNN) on top to provide different types of predictions, e.g. of 
the probability of distant metastasis over time
\cite{Kazmierski_Haibe-Kains_2021}, or the PFS~\cite{Lv_Zhou_Peng_Peng_Lin_Wu_Xu_Lu_2023,Aswathi_Rizkallah_Frecon_Bailly_Bodet-Milin_Casasnovas_Gouill_Kraeber-Bodéré_Carlier_Mateus_2023}.
%, respectively). %Developper cette graph structure et le GCN associé  %PFS, proability of distant metastasis over time
Aswathi et al. \cite{Aswathi_Rizkallah_Frecon_Bailly_Bodet-Milin_Casasnovas_Gouill_Kraeber-Bodéré_Carlier_Mateus_2023} %proposed a method where 
exploit only imaging descriptors taken from multiple lesions,
%are exploited
while \cite{Lv_Zhou_Peng_Peng_Lin_Wu_Xu_Lu_2023} and \cite{Kazmierski_Haibe-Kains_2021} consider a naive late fusion to incorporate clinical information, i.e. the clinical features are concatenated with %the imaging ones 
imaging descriptors just before the prediction computation at the last fully connected layer. However, %none of the above approaches study alternative methods for the fusion of data arising from both multiple lesions and multiple data modalities.
given the naive fusion's simplicity, alternative approaches are needed to study the fusion of multiple lesions and heterogenous data modalities.
% \mdf{DM: If other papers rely mostly on late fusion we might want to compare against }
% \textit{Oui mais ils ne sont pas sur les mêmes tâches (à part Aswathi contre qui on a testé), donc il faudrait qu'on réimplémente tout :/}
% \mdf{DM:je pensais a une concatenation vs la crossatention: gardant le reste du modèle identique} \mr{Je pensais à la même chose que Diana, un MLP qui prend en entrée : une moyenne de features de tous les lesions/ou les features de la lesion la plus fixante + données cliniques et output la PFS deux ans?? } \textit{Je l'ai testé ça, c'est MLP image + clinique dans les résultats !}

%\subsection{Multimodal data fusion}
%\mdf{Mira : DIANA, can you check this part? }
%\mdf{Mira : Et dans l'autre sous section parler vers comment incorporer les données cliniques/multimodal fusion et citer les papiers qui font late fusion / early fusion et se situer par rapport à ceux là.}

Beyond PET imaging and cancer risk prediction, there has been an increasing interest in fusing the information from multiple modalities to perform better-informed predictions. As discussed by Baltrušaitis et al.~\cite{Baltrušaitis_Ahuja_Morency_2019}, there are multiple ways of fusing multimodal data, e.g. %have been proposed in the literature, 
%such as 
the classical: early, late and hybrid fusion approaches, kernel-based methods, graphical models and some neural networks. However, %for the time being, 
none %has been able to establish itself as 
%a consensual, state-of-the art method, 
is today consensual
%, especially when dealing with 
for dealing with heterogeneous medical data.
%For reference, the sixth part of the paper \cite{Baltrušaitis_Ahuja_Morency_2019} discuss a variety of ways to do this fusion. %Est-ce que ça suffit de juste citer un article de review?

%A way that has been used to fuse multiple modalities of data for bio-medical applications is cross-attention. 
Recently, cross-attention modules have been explored to fuse multiple modalities in bio-medical applications. For instance, Mo et al. \cite{Mo_Cai_Lin_Tong_Chen_Wang_Hu_Iwamoto_Han_Chen_2022} implemented %this method
a cross-attention strategy to fuse the information from two MRI imaging modalities for a segmentation task.
% to integrate two graphs by computing the cosine similarity between all the nodes from each graph, with each graph representing an image for a segmentation task. 
Chen et al. \cite{Chen_Liu_He_Du_2023} computed a cross-attention based on transformers %models 
\cite{Vaswani_Shazeer_Parmar_Uszkoreit_Jones_Gomez_Kaiser_Polosukhin_2017} to %perform image registration between 
register two imaging modalities, 
by considering different modalities for query than for the keys/values. 
Finally, targeting heterogenous data, Bhalodia et al. \cite{Bhalodia_Hatamizadeh_Tam_Xu_Wang_Turkbey_Xu_2021} used %it 
cross-attention for pneumonia localization by computing cosine similarities between images and text embeddings. 
%
%In the same way, we took inspiration from \cite{Xie_Li_Zhao_Pan_Wang_2023}, who proposed a way to fuse vectorial and graph data (as in our problem) by using a cross-attention module as in \cite{Vaswani_Shazeer_Parmar_Uszkoreit_Jones_Gomez_Kaiser_Polosukhin_2017} (initially applied to the problem of open relation extraction). 
Beyond the medical domain but relying on graphs, Xie et al.~\cite{Xie_Li_Zhao_Pan_Wang_2023} proposed to fuse vectorial and graph data with cross-attention modules for open relation extraction in text analysis.

%We combined this approach with the one from \cite{Chen_Liu_He_Du_2023}, who used the same cross-attention module to fuse data by considering different information as query and key/value. 
In this work, we build a multi-lesion graph to capture image and structural properties~\cite{Lv_Zhou_Peng_Peng_Lin_Wu_Xu_Lu_2023,Kazmierski_Haibe-Kains_2021,Aswathi_Rizkallah_Frecon_Bailly_Bodet-Milin_Casasnovas_Gouill_Kraeber-Bodéré_Carlier_Mateus_2023}. In addition, we
take inspiration from \cite{Chen_Liu_He_Du_2023} and \cite{Xie_Li_Zhao_Pan_Wang_2023}
to propose a cross-attention method between the image lesion graph and clinical data. 
%by considering different information as query and key/value. 
%
%We combine these two ideas to apply them to the problem of the prediction of high-risk DLBCL patients. %Thus, our method brings together the information both from the multiple lesions of the patient and from the different available modalities to improve the identification of high-risk patients in the DLBCL context.
%We rely on 
The proposed model addresses the identification of high-risk patients in DLBCL.

%en fait le pb de survie n'a rien à voir avec la multimodalité, pourquoi on se restreint à ça ? À la limite on peut plutôt parler du pb des graphes (mais je n'ai pas bcp creusé ça, dur de faire une étude exhaustive) ? Ou du domaine médical (même si je ne sais pas s'il y a des pbs associés, pareil)

%best of both worlds ? : multimodal contrastive learning

%An Ensemble Approach for Patient Prognosis of Head and Neck Tumor Using Multimodal Data : Features are extracted from fused CT and PET scans using the CNN network and concatenated with the EHR features. Then, the output is passed to the FC layers before MTLR. Lastly, risk scores from MTLR and CoxPH models are averaged to get the final risk predictions.

%Predicting the Survival of Cancer Patients With Multimodal Graph Neural Network : concatène les sorties de chaque layer avant FC finale

%Multi-Modal Graph Learning for Disease Prediction : plein de trucs dont cross-att, à creuser

%Est-ce qu'on est vraiment censés être exhaustifs ? (au moins mentionner les derniers/meilleurs)

%Plein, mentionner en une fois tous ceux qui ont des concaténations naïves, plus développer un peu plus ceux qui utilisent des choses plus subtiles ? 
%Dire les articles qui font de la cross-attention comme nous (et trouver ce qu'on fait en plus)

%
\section{Method}
\label{section_meth}

\subsubsection{Problem statement}
\begin{figure}[h]    
\centering\includegraphics[width=0.93\textwidth]{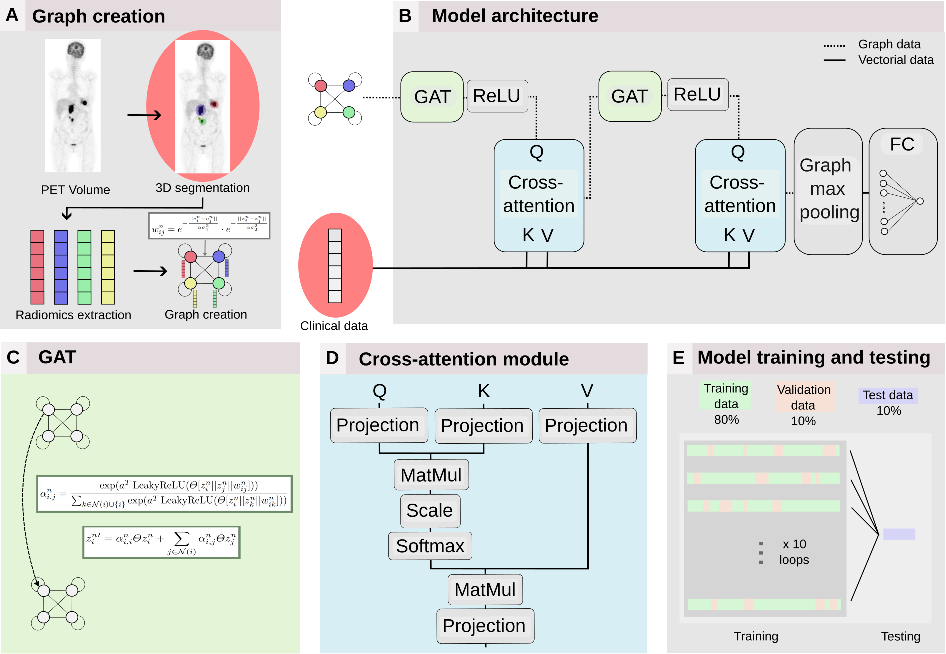}
\caption{Method overview: (A) patient-level graph with imaging information from every lesion, (B) model architecture, propagating the information from the multiple nodes (with the GATv2) and fusing it with the clinical data by the cross-attention block, (C) explanation of the GATv2, (D)~the cross-attention mechanism, (E) %dm data organisation for the training and testing of the model. 
training and testing schemes.
The red circles indicate the patient's information provided in the dataset.} \label{full_method}
\end{figure}

Let a DLBCL clinical exam before treatment be composed of a full-body PET image acquired on a patient, and a set of tabular clinical indicators. 
%dm (see Supplementary material \ref{supp_clin}) at the time of the exam. 
Our goal is to perform a PFS 2-year classification, intended to predict whether the disease of a patient will progress within two years after the beginning of the treatment. This indicator helps to identify high-risk patients (more likely to progress). In this context, we propose a learning framework (c.f. Fig.~\ref{full_method}), taking as input clinical tabular data and a full-body 3D PET image with 3D segmentation %manual 3D \dm{polygonal? bounding box} segmentation 
of the lesions, trained to predict a probability of 2-year PFS. 

%Our framework is composed of $4$ %a succession of $4$ types of 
%modules. 
%The first module consists of a lesion graph generation, 
%where we define both the underlying structure of the lesions and the feature vector assigned to each of them. 
First, we design a \textit{lesion graph} to simultaneously represent the image features of individual lesions and their spatial distribution. Then,
a %A raph neural network (GNN) 
%\dm{two-layer} 
GNN is built on the top of the constructed graph, composed of
%, composed of two consecutive layers. 
%each layer consists of 
i) \textit{graph attention} %model 
modules that learns a latent representation from multiple lesions; and ii) a \textit{cross-modal} fusion blocks integrating clinical data.
%The output of the second layer enters the prediction module. 
A final \textit{prediction} module aggregates the fused information into a classification score.
%In the following, we provide a detailed description of %each of the four 
%the modules.
%We want to learn a binary prediction of whether the disease of a patient progressed within two years after the beginning of the treatment (2-years PFS). For this, we take as inputs a full-body 3D PET image on which the lesions are segmented and clinical tabular data. The proposed framework is made of four main steps, described in Figure \ref{full_method}: first is the creation of a lesion graph, followed by the learning phase composed of the succession of a GATv2 model and a cross-attention layer, and finally the prediction module. These steps are explained in more detail in this section. %Détailler un poil plus ? (oui : on ne sait pas ce que sont ces 4 steps avec l'image)

%\textit{Décrire brièvement la structure du modèle qui arrive ensuite}

\subsubsection{Lesion Graph construction} %voir ce que je renvoie à Aswathi et ce que je redis (il doit être suffisant de lire notre article)
\label{graph_crea}
%Dire qq part qu'on veut étudier la position relative des nodes, les relations qu'elles ont entre elles, etc
%\textit{Description de la façon dont est construit un graphe avec 1 node=1 lésion, les nodes features correspondant aux radiomiques et aux données classiques, le choix des edges (fully connected avec self-loops) et l'équation régissant les edge weights} %Mettre en sup mat les tableaux des features rad et class
The first step of our framework is the creation of a fully connected graph $\mathcal{G}^{(n)} = \{\mathcal{V}^{(n)}, \mathcal{E}^{(n)} \}$ to group the information from the %multiple 
$L^{(n)}$ lesions present on the PET scan of the $n^{th}$ patient. We construct this graph as in \cite{Aswathi_Rizkallah_Frecon_Bailly_Bodet-Milin_Casasnovas_Gouill_Kraeber-Bodéré_Carlier_Mateus_2023}: each node $v^{(n)}_i \in \mathcal{V}^{(n)}$ corresponds to a single lesion, and is associated with a feature vector $\mathbf{z}^{(n)}_i\in\mathbb{R}^{D_{\mathrm{features}}}$. This vector contains both classical intensity-based and radiomics features\footnote{
%We define as classical those conventional quantitative measurements on the segmented lesion describing the intensity distribution of the voxels without taking into account their spatial relationships. Instead, radiomics features are quantitative features describing the 3D structure of the lesion, such as shape or second-order features that reveal the inter-relationship among voxels.
Here, classical features are quantitative measurements on the segmented lesion describing the intensity distribution of the voxels. Radiomics features instead  describe the 3D structure of the lesion, such as shape, or second-order features that reveal the inter-relationship among voxels.}(c.f Table 1 in the Supp. material). In the following, we denote by 
%$\mathbf{Z}^{(n)} \in \mathbb{R}^{|\mathcal{V}^{(n)}| \times D_{\mathrm{features}}}$ 
$\mathbf{Z}^{(n)} \in \mathbb{R}^{L^{(n)} \times D_{\mathrm{features}}}$ 
the matrix concatenating all nodes' features $\mathbf{z}_i^{(n)}$, with $D_{\mathrm{features}}$ 
%being the sum of the number of 
the dimension of the %feature 
vector including
classical and radiomics features.

Edges $e^{(n)}_{ij}$ are drawn between every pair of nodes $v^{(n)}_i$ and $v^{(n)}_j$, including self-loops.
Weights $w^{(n)}_{ij}$ are assigned to each edge to favor message passing between closer and more similar lesions. The values of $w^{(n)}_{ij}$ are defined based on the distances between both the feature vectors $\mathbf{z}^{(n)}_i$ and the lesions centroids $\mathbf{p}_i^{(n)}$: %justifier ces choix ? 
\begin{equation}
    w^{(n)}_{ij} = \exp\left({- \frac{||\mathbf{p}^{(n)}_i - \mathbf{p}^{(n)}_j||_2}{\gamma\sigma_1^2}} 
    \right)
    \cdot 
    \exp\left({- \frac{||\mathbf{z}^{(n)}_i - \mathbf{z}^{(n)}_j||_2}{\gamma\sigma_2^2}}
    \right),
    \label{eq:weights}
\end{equation}
where $||.||_2$ stands for the L2 %vector 
norm; $\sigma_1$, $\sigma_2$ denote the population-level standard deviations of the centroid and the feature distances, respectively;
%distances between the centroids' coordinates and between the feature vectors, respectively. 
and $\gamma$ is a hyper-parameter tuned to find the best edge weight distribution for our task.%the PFS classification task.
%N ensemble des nodes, n un node et z sa feature
%E ensemble des edges, e un edge et w sa feature

% The first step of our framework is the creation of a fully connected graph to group the information from the multiple lesions present on the PET scan of the patients. Each lesion $i$ of the $n^{th}$ patient corresponds to a node of the graph, which is associated with its node features $Z^n_i$. The node features are both radiomics extracted from this lesion on the PET image and classical data (cf Table ..)

% Edges em,n are drawn between every pair of nodes m, n $\in$ Ni , including a self loop

% A graph is defined by a set of nodes .., linked by a set of edges .. . Both the nodes and the edges may be linked to a set of attributes %à voir

% In our framework, we define a graph as follows: each node corresponds to a lesion, and thus contains as node features radiomics extracted from this lesion on the PET image. %Préciser lesquelles    These radiomics are .. % ajouter les données classical !!
% The graph is fully connected with self-loops, and these edges are weighted depending on both the distance between the features of the two nodes and the Euclidean distance between the 3D centroid coordinates of each lesion $c_\cdot^n$: %justifier ces choix ?

% \begin{equation}
%     w_{ij} = e^{- \frac{||c^n_i - c^n_j||}{a\sigma_1^2}} \cdot e^{- \frac{||z^n_i - z^n_j||}{a\sigma_2^2}}
% \end{equation}
% %Citer Aswathi là ? Remplacer mm ce paragraphe par un renvoi a son article ?

\subsubsection{Multi-Lesion Representation Learning} %Vérifier cohérence des notations math.
%\textit{Expliquer le principe du GAT (et son intérêt ici ?)}
To study the relations between the lesions and to pool their information, we define a GNN over our lesion graph. We rely on the GATv2 convolution layer \cite{brody2022how} for its capacity to adapt the neighbors' attention weights independently for each node. In our context, the attention scheme of GATv2 implies that the feature vector of each lesion is updated based on information propagated from the most relevant neighboring lesions only. 
%\mdf{->"its capacity to adapt the neighbour attention weights independently for each node". In our context, the attention scheme of GATv2 implies that the feature vector of each lesion is updated based on information propagated from the most relevant neighbouring lesions only.
% We could go futher but not sure it is needed
%In more detail, the feature vector of lesion $i$, is updated with a weighted sum of projected self and neighbour features $j\in {i \cup Nei(i)}$, with learnt weights modulated by an attention coefficient dependent on each feature pair $zi$ and $zj$ and the edge weight $w_ij$
%}
%\sout{GATv2 usually outperforms its predecessor GAT because of the way it dynamically compute the attention weights of a node depending on its own representation. }
We implement the \texttt{torch\_geometric} version of this operator,
%\footnote{\url{https://pytorch-geometric.readthedocs.io/en/latest/generated/torch_geometric.nn.conv.GATv2Conv.html}}, 
which takes into account edge weights by computing the attention coefficients $\alpha_{i,j}$ as follows: 

\begin{equation}
    \alpha_{i,j}^{(n)} = \frac{\exp(\mathbf{a}^T\LeakyReLU(\mathbf{\Theta} [\mathbf{z}^{(n)}_i||\mathbf{z}^{(n)}_j||w^{(n)}_{ij}]))}{\sum_{k \in \mathcal{N}(i)\cup\{i\}}\exp(\mathbf{a}^T\LeakyReLU(\mathbf{\Theta} [\mathbf{z}^{(n)}_i||\mathbf{z}^{(n)}_k||w^{(n)}_{ik}]))},
\end{equation}
with $\mathbf{a}$ and $\mathbf{\Theta}$ learned parameter matrices, $\cdot||\cdot$ the concatenation operation and $\mathcal{N}(i)$ the neighboring nodes of $v_i^{(n)}$. 
The features assigned to each lesion (i.e. node in the graph) are updated as: %dm follows:
\begin{equation}
    \mathbf{z}_{i_{\mathrm{GAT}}}^{(n)} = \alpha_{i,i}^{(n)} \mathbf{\Theta} \mathbf{z}_i^{(n)} + \sum_{j \in \mathcal{N}(i)} \alpha_{i,j}^{(n)} \mathbf{\Theta} \mathbf{z}_j^{(n)}.
\end{equation}

The $D_{\mathrm{GAT}}$ dimension of %each 
the node's representation $\mathbf{z}_{i_{\mathrm{GAT}}}^{(n)}$ at the output of a GATv2 block is determined by a grid search, as %well as the 
 is also the dropout probability
%probability of the dropout 
applied to this module. Finally, the updated lesion representations are passed through a ReLU activation. The resultant 
{$L^{(n)}\times D_{\mathrm{GAT}}$}
feature matrix %is defined as a column-wise 
is a concatenation of the lesions feature vectors: $\mathbf{Z}_{\mathrm{GAT}}^{(n)} = \ReLU([ \mathbf{z}^{(n)}_{{1}_{\mathrm{GAT}}}
%, \ldots, 
||\ldots ||
\mathbf{z}^{(n)}_{{L^{(n)}}_{\mathrm{GAT}}} ]^\top)$. %where $L=\mathcal{V}^{(n)}$ is the total number of lesions in the graph, preserved after the GATv2 layer.
%which preserves the number of lesions $L$.
%As an activation layer, our GATv2 is followed by a ReLU operation.
%\mdf{MIRA: %What about dropout? non linear activation layers? 
%the input output dimensions of the hidden channels, the latent space dimensions ? 
%why GATv2 why not GAT? 
%notations : bold lower case for vectors, bold upper case for matrices, normal lower case for values. 
%A small phrase explaining how GATv2 do information propagation between different lesions taking into acccount the weighs? How can this be interesting for our problem? }

\subsubsection{Multimodal Multi-lesion Cross-Attention}
%\textit{Expliquer le principe du module de cross-attention et comment on l'utilise}
%\mdf{MIRA: Je vais retravailler cette partie.  C'est notre contribution principale. ON va la mettre en avant. COmmencer par l'intêret, on a des données cliniques, on a des données de différentes lésions, comment peut on voir la contribution de chaque feature des lesions sur chacune des features de cliniques, c ca?? } 
%\textit{alors oui je dirais que c'est ça, même si pour moi ça aurait plus de sens de faire l'inverse (surtout qu'on recupère un graphe en sortie) ^^'} \mdf{je comprend. Le Z GAT c une matrice ? ou un vecteur? puisque dans la partie juste avant zigat c'était un vecteur. Ici c devenu en majuscule, donc matrice?  Faut  définir le ZGAT par rapport à ziGAT} \textit{alors le Z c'est la matrice concaténant les petits z de chaque node normalement ligne 237 du code tu devrais en trouver la definition} \textit{et si tu veux parler par le chat ça sera plus facile si tu veux ;)}

%At this stage, for the $n^{th}$ patient, we have a set of updated feature vectors assigned to the nodes of the lesion graph $\mathbf{Z}_{\mathrm{GAT}}^{(n)}$.  
We aim now at projecting the updated node features $\mathbf{Z}_{\mathrm{GAT}}^{(n)}$, into a more representative space by integrating the %prior 
clinical knowledge of the patient $n$ represented by a vector 
%$\mathbf{c}^{(n)} \in \mathcal{R}^{A \times D_{\mathrm{clin}}}$ \pb{$\mapsto$ What is A?$\mapsto$ 
$\mathbf{c}^{(n)} \in \mathbb{R}^{D_{\mathrm{clin}}}$; note that there is a vector $\mathbf{c}^{(n)}$ per patient (and not per lesion). 
%Once information from all the nodes was fused  we integrate it with the vector of clinical data $c^{(n)}$. This clinical data is fused with the graph output of the GATv2 layer by a cross-attention module. 
For this purpose, we take advantage of the self-attention module proposed in \cite{Vaswani_Shazeer_Parmar_Uszkoreit_Jones_Gomez_Kaiser_Polosukhin_2017} adapted to the cross-modal case. 
The module takes as input a query vector $\mathbf{Q}$ and a key/value pair of vectors $\mathbf{K}$ and $\mathbf{V}$ and outputs a weighted sum of the values, where the weight assigned to each value is computed from a compatibility function (\textit{i.e.} a scalar product) of the query with the corresponding key (normalized by the key dimension $d_k$). %$d_k=\dm{D_{\mathrm{clin}}}$). %of the key vector). 
By defining $\mathbf{Q} = \mathbf{Z}_{\mathrm{GAT}}^{(n)}$ and $\mathbf{K} = \mathbf{V} = \mathbf{c}^{(n)}$, the signals assigned to each lesion are %hence 
updated with the information procured by the clinical data:%both the clinical data and the \dm{image features:}
%features of the lesion latent space learned by the previous graph attention module:
%Pb : d'où Q et K ont la mm dim ? Laissé que k mais pas de preuve, j'arrive pas à trouver le code correspondant même sur le Github
%Poids W sur les attentions !! (à expliquer pq on les a (et checker si on les a vraiment en n'utilisant qu'une seule head)
%\mr{\begin{equation}
\begin{align}
   \mathbf{Z}_{\mathrm{CrossAtt}}^{(n)} &= \textnormal{CrossAtt}(\mathbf{Z}_{\mathrm{GAT}}^{(n)},\mathbf{c}^{(n)},\mathbf{c}^{(n)}) \nonumber \\
&=%\Big(
\softmax\bigg(\frac{\mathbf{Z}_{\mathrm{GAT}}^{(n)}\mathbf{W}^Q(\mathbf{c}^{(n)}\mathbf{W}^K)^T}{\sqrt{d_k}}\bigg)\mathbf{c}^{(n)}\mathbf{W}^V.%\Big)\mathbf{W}^O.
\end{align}

% \begin{align}
%    \textnormal{CrossAtt}(\mathbf{Q},\mathbf{K},\mathbf{V}) =
% \softmax\bigg(\frac{\mathbf{Q}\mathbf{W}^Q(\mathbf{K}\mathbf{W}^K)^T}{\sqrt{d_k}}\bigg)\mathbf{V}\mathbf{W}^V.%\Big)\mathbf{W}^O.
% \end{align}
%\end{equation}}

We optimize during training the latent representations of $\mathbf{Q}$, $\mathbf{K}$ and $\mathbf{V}$ and the cross attention output via three learnable %parameter 
matrices 
$\mathbf{W}^Q\in\mathbb{R}^{D_{\rm GAT}\times D_{\mathrm{clin}}}$, $\mathbf{W}^K\in\mathbb{R}^{1\times D_{\mathrm{clin}}}$ and 
$\mathbf{W}^V\in\mathbb{R}^{1\times D_{\mathrm{\rm GAT}}}$.
%with size respectively of $D_{\mathrm{GAT}}\times D_{\mathrm{clin}}$, 
%$D_{\mathrm{clin}} \times 1$ and $D_{\mathrm{GAT}} \times 1$. 
%\pb{$1\times D_{\mathrm{clin}}$ and $1\times D_{\mathrm{GAT}}$}. 
% and $\mathbf{W}^O$. % with size.
The result of the cross-attention operation is a matrix $\mathbf{Z}_{\mathrm{CrossAtt}}^{(n)}$ of same size as~
$\mathbf{Q}$ ($L^{(n)} \times D_{\mathrm{CrossAtt}}=L ^{(n)}\times D_{\mathrm{GAT}}$). 
%$\mathbf{Q}$ ($|\mathcal{V}^{(n)}| \times D_{\mathrm{CrossAtt}}=|\mathcal{V}^{(n)}| \times D_{\mathrm{GAT}}$). 
Intuitively, matrices $\mathbf{W}^Q$ and $\mathbf{W}^K$ project the multi-lesion image data and the clinical vector on to a common space, before computing their compatibility. The softmax output, of size ($L^{(n)}\times D_{\rm clin}$), provides the attention values that each lesion should give to the entries of the clinical vector. Finally, the attention scores are multiplied with the clinical data vector, lifted to the $D_{\rm GAT}$ dimension by $\mathbf{W}^V$.
%We get back to the node features by splitting the matrix \dm{ $\mathbf{Z}_{\mathrm{CrossAtt}}^{(n)}$} along the $L$ nodes dimension. 
The updated individual node features correspond to the rows of  $\mathbf{Z}_{\mathrm{CrossAtt}}^{(n)}$.
%Préciser comment on lie les dimensions des deux
%voir s'il y a besoin d'expliquer qqch par rapport à l'aspect graph
% \mdf{MIRA : Cross attention entre les features d'un noeud dans l'espace latent et les indicateurs cliniques? 
% L'attention est calculée par noeud du graphe ou pas? 
% What is Q, K, V in your case?  
% Les dimensions de la sortie de ta cross attention ca vaudra le coup d'en parler même si tu donne pas des valeurs précises les mettre et les ajouter sur ta figure? 
% À la sortie de la Cross attention on a des valeurs d'attention et aussi un update des features sur chaque noeud, non ? 
% Je suis un peu perdue sur la concatenation des z, je comprend pas comment tu les concatène. Need more details..  
% expliciter le calcul fait sur la matrice des features $\mathbf{Z}$ avec l'embedding fait avec le GAT, ensuite la matrice résultante en utilisant les cross-attentions. donc chaque noeud : $\mathbf{z}_i$ updaté devient quoi? }
The multi-lesion and cross attention modules are repeated for  two layers. After the second layer, we end up with 
$\mathbf{Z}_{\rm CrossAtt}^{(n)}{}'\in \mathbb{R}^{L^{(n)} \times D_{\mathrm{GAT}}}$.

\subsubsection{Prediction}
A max pooling on $\mathbf{Z}_{\mathrm{CAtt}}^{(n)}{}'$, across the node dimension, 
%is applied to the output of the second cross-attention block 
%$\mathbf{Z}_{\mathrm{CrossAtt}}^{(n)}{}'\in \mathbb{R}^{L \times D_{\mathrm{GAT}}}$, 
%in order to get 
resumes the graph features to a $D_{\mathrm{GAT}}$-dimensional vector, allowing us to handle patients with different numbers of lesions.
%. Finally, the latter enters 
The pooled vector is given to a linear layer with a sigmoid activation function to make a prediction of the 2-year PFS for a given patient. 
The learning is controlled by a weighted binary cross-entropy loss function, where weights compensate for the class imbalance
%. As the classes are %quite unbalanced
%\dm{imbalanced} 
(ratio of positive to negative samples $\sim 1:5$).
%, we introduce in our loss function a class weight proportional to the ratio of training data of each class.

%\textit{Dire que les étapes précédentes sont répétées deux fois et qu'on utilise ensuite un graph pooling et une couche complètement connectée pour la prédiction binaire}

% \mdf{je trouve celà un peu vague.. pourquoi pas expliciter la matrice de features des noeuds (de dimension V x Dfeatures, comment elle varie dans les différentes couches, et comment le pooling s'éffectue selon la dimension des noeuds pour se retrouver avec seulement un noeud et son vecteur associé.}
%expliciter max-pooling ? (sur quelle dim, etc)
%confirmer le max pooling avec gsearch
\section{Experiments}

\subsubsection{Dataset}
The proposed method was evaluated on the prospective GAINED study (NCT 01659099) \cite{gouill2021} which enrolled $670$ newly diagnosed and untreated DLBCL patients. In order to perform our binary prediction of the 2-year PFS, we removed the patients who were censored before this time, which left us with $583$ samples. Among these patients, 101 were deemed as positive for the PFS because of a progression or a relapse of the disease within two years, while 12 were positive because of death without progression of the disease. In this dataset, are assigned to each patient a PET image at the beginning of the protocol as well as clinical indicators such as age, ECOG scale%aaIPI (age-adjusted International Prognostic Index)
, Ann Arbor stage or number of extranodal sites (full list %of the clinical data considered 
is presented in Supp. material). The lesion detection on the PET images is done manually by a clinician and the segmentation is performed using a majority vote between three usual lesion segmentation methods: 
i) a K-means clustering %algorithm
($K=2$), %clusters, 
ii) %considering as lesion 
a thresholding that retains only voxels with intensity values larger than 41\% of the maximum intensity, %voxel, 
and 
iii) %taking only voxels whose value in the 
a second thresholding to keep voxels whose normalized SUV (Standard Uptake Value) is more than 2.5.
The imaging and clinical features are both standardised by removing the median of the training data and scaling the whole dataset according to the quartile range:
%\begin{equation}
$
    Scaled~value = \frac{Original~value - training~median}{training~interquartile~range}.
$
%\end{equation}
The distance between the centroid of the lesions (Eq.~\ref{eq:weights})
%, used for the lesion graph creation, 
is  standardized in a similar %same 
way, but considering the mean and quartiles of the lesions' centroids individually for each patient.
%of only the considered patient.

%Valeur NCT ... que pour la phase 3 de GAINED ? Préciser cette histoire de phase ?
%Multicentrique
%prospective study
%other data? (sex proportion, age, prop labels
%clinical (which ones ? only those that we use or all of them?) (mettre toutes celles qu'on utilise au moins?) +PET images (caracs ?)
%enlevé patients censored until 583
%distribution du nb de lésions
%Si pas expliqué censure : dire juste patients who left the study before this time
%Expliquer que notre grid search a pas été faite comme sur les baselines

\subsubsection{Comparison to baseline models}
%\textit{Présenter les baseline models (leurs entrées et les modèles utilisés)} %Mettre en supp mat les configs retenues des modèles

%The proposed 
Our model was compared to six other baseline models %also performing a 2-year PFS binary classification
performing the same task: %three Multi-Layer Perceptrons (MLPs) with different inputs; one Multiple Instance Learning (MIL) approach %directly 
%applied to a bag of lesions; and a Graph Neural Network (GNN) baseline:%. Below, we detail each approach: % and delivering as output, a \mdf{expliciter the output est-ce une probabilité pour dire si le disease a progressé ou pas}. 
%% Mira : j'ai retiré ici
\begin{itemize}
    \item \textbf{MLP clinical}: An MLP whose only input is the vector of clinical data of a patient. The model comprises two linear layers with ReLU activations and a 1-dim linear output layer with sigmoid activation. The two intermediate layers have the same dimension, in practice chosen via a grid search.  %Préciser MLP ?  
    \item \textbf{MLP image}: An MLP with the same configuration but taking as input the imaging data. We compute the input image vector as the average of the feature vectors from individual lesions to handle the variable lesion number across patients. For each lesion we extract features as in Sec. \ref{graph_crea}.
    \item \textbf{MLP clinical+image}: An MLP, with the same configuration as the previous ones, but taking as input the concatenation of both the clinical and imaging data (\textit{i.e.} the input image vector 
    as for the MLP image).
    %as defined above). %à préciser/finir 

    %\item A MLP, with the same configuration as the previous one, which takes as input both the clinical data of the patient and graph imaging data. This graph imaging data corresponds to graph features (number of node, number of edges and maximum distance between the nodes) extracted from a 3-nearest neighbors graph created as presented before (excepted for the edges definition). %à préciser/finir 
    %\mdf{Je ne sais pas si c'est interressant ca. EN VRAI en y repensant, le fait de faire un 3-NN avant de calculer la distance, on loupe la dispersion (un indicateur important), qui est la distance max entre nimporte quel deux noeuds du graphe.}
    
    \item \textbf{MIL image}: A MIL approach taking as input the imaging features from the $L^{(n)}$ %multiple 
    lesions of a patient, applies a one-layer MLP followed by a ReLU on each lesion's feature vector, aggregates the results by a maximum operation and projects it linearly (with a sigmoid activation) to get the prediction.
    \item \textbf{GraphConv image}: A GraphConv model \cite{Morris_Ritzert_Fey_Hamilton_Lenssen_Rattan_Grohe_2019}, taking as input a lesion graph as %defined 
    in Sec.~\ref{section_meth}, but using a %classical 
    graph convolution aggregation function, see Eq.~\ref{eq:graphconv}.
    The model is composed of two GraphConv layers, the first having an output dimension determined by grid search, and the second with an output size of 1. The first layer has a ReLU activation, and the second is followed by a max pooling operation and a sigmoid activation to predict the PFS.
    \begin{equation}
        \mathbf{z}_{i_{\mathrm{GraphConv}}}^{(n)} = \mathbf{W}_1 \mathbf{z}_i^{(n)} + \mathbf{W}_2 \bigg(\sum_{j \in \mathcal{N}(i)} w_{ij}^{(n)}\mathbf{z}_j^{(n)}\bigg).
        \label{eq:graphconv}
    \end{equation}
      %of the patient.
    %\item \textbf{GATv2 image}: A GATv2 model (without the cross-attention module) which takes as input the same imaging graph and has the same structure as the GraphConv model.
\end{itemize}

%Préciser la configuration un peu plus précise des modèles (2 couches, etc)

\subsubsection{Ablation study}
In order to prove the interest of each module in our framework
%element of our model, 
we also do two ablation studies. First, we implement our model with GraphConv layers replacing the GATv2 layers to study the impact of the learned attention weights between the lesions. Then, we replace the cross-attention layers by a simple concatenation 
$[\mathbf{z}^{(n)}_{i_{GAT}}||\mathbf{c}^{(n)}]$
%to see if the learning of the fusion 
to verify if the proposed learnable fusion
between the two modalities improves the performance of the model. %Toward the same goal, we implement a model based on two GATv2 layers with clinical data concatenated to the last linear layer used for the prediction. %%%%%%%%%%%%%%%%%%%%%%%%%%%%%%%%%%%%%%%%%%%%%%%%%%%%%%%%%%%%%%%%%%%%%%%%%%%%%%%%%%%%%%%%%%%%%%%

\subsubsection{Experimental setup}
%\textit{Parler du choix des paramètres et de la gsearch, dire qu'on utilise le test de student pour valider la pertinence de nos résultats et expliquer que ça a été codé en Python avec Torch et torch\_geometric}
%\textit{Présenter notre framework d'entraînement/validation/test}
We strictly divide the %data from the 
583 patients in three distinct sets of training (80\%), validation (10\%) and test (10\%). %, ensuring there is no patient overlap between the different sets. %But it seemed obvious to me with what was said...
Test results are reported for the model with the best validation ROC AUC.
To evaluate our model, the %previous process 
split is repeated ten times as follows: a single %set of test 
test set is left out from all the loops, and at each loop the remaining data is randomly split into training and validation sets, while ensuring that the ratio of positive patients is the same in all the sets. 
%To get more informative results, for each loop we do five runs of validation and test: on each run we test the model on all the positive data and on 1/5 of the negative data to enforce a test on a balanced dataset. 
Furthermore, to ensure the scores are computed on balanced sets, we repeat the validation and test phases five times: for each run we build a balanced set with all the available positive data and 1/5 of the negative data, randomly sampled from the validation and test sets respectively. The resulting metrics are then averaged to get the final validation or test results. A grid search (c.f. Table~3 in Supp. material) is performed on the learning rate, the hidden channel size and, for the GNN, the parameter $\gamma$ (used in the lesion graphs construction) to find the model configuration that grants the best validation ROC AUC. %The chosen configurations, whose results are presented in this paper, are detailed in Supplementary material 3 along with all tested configurations. 
Furthermore, in order to validate the statistical significance of our results, we use a t-test %\cite{Cohen_1988} 
to compare the results of our model against the baselines. The whole framework has been coded in Python with \texttt{PyTorch} and \texttt{torch\_geometric} modules. %Checker qu'il n'y a que ça

\section{Results}

% \textit{Mettre tableau des résultats des baselines (avec les p-values indiquant que notre modèle est toujours significativement meilleur)}

% \textit{Discuter ce tableau : expliquer que les données cliniques apportent plus que les données images, qu'il est nécessaire d'utiliser toutes les lésions (MIL$>$MLP\_image) et que notre méthode permet d'intégrer de façon efficace toutes ces informations}

% \textit{Rappeler la difficulté/variabilité de la segmentation et que ça peut expliquer les mauvais résultats liés aux données images}

% \textit{Présenter les perspectives : utiliser des graphes de sous-régions de lésions pour prendre en compte ces soucis de segmentation, tester sur d'autres jeux de données pour montrer la généralisabilité du modèle, explorer des fonctions de coût liées à la survie pour affiner nos prédictions}

\subsubsection{Quantitative results} 
We report in Table 1 %\ref{table_results} 
the results of our comparative study.
Our experiments reveal that models based on clinical data perform better than models using imaging data only.
Furthermore, for models based on imaging data, considering the lesions individually (as the nodes of a graph or a bag of nodes in the MIL) seems to improve the predictions compared to averaging the feature vectors. %from all the lesions
Also, using a graph improves over the bag of lesions/MIL approach. 
Finally, the proposed framework  performs significantly better than all the other models (p-value < 0.005), showing it efficiently fuses the information from multiple lesions and from the two considered modalities.

For the \textbf{ablation} studies, replacing the cross-attention layers by a simple concatenation %has a big impact on the performances of the model 
results in a big performance drop (test ROC AUC of $0.59\pm0.06$ against $0.72\pm0.03$ initially)%, as well as doing a late fusion between the modalities (test ROC AUC of $0.54\pm0.06$)% against $0.72\pm0.03$)
, proving the benefit of our multimodal data fusion method. However, replacing the GATv2 layers with GraphConv layers does not significantly affect the performances (test ROC AUC of $0.71\pm0.04$). %which is a bit unexpected.

 %The obtained results may seem weak but are expected for a treatment response prediction task, which is a complex problem with few data and no known precise imaging predictor. 
 The better performance of clinical-based models
 %models based on clinical data 
 compared to those based on imaging %data 
 can be partially explained by the selection of a subset of clinical variables known for being predictive \cite{gouill2021}. Another aspect influencing the image-based models is the high complexity %and thus high intra-operator variability 
of the lesion segmentation task for DLBCL patients given that lesions tend to superpose and have diffuse contours. Nonetheless, we argue that an efficient integration of both kinds of data, and all the lesions, as proposed here, should allow for a better assessment of  the patient's state.
\vspace{-0.5cm}
\begin{table}[h]
\caption{Test ROC AUC of the considered models (best performance in bold), with the p-value comparing the results to those of the cross-attention model.  %image* refers to using the mean of the features of the lesions as our imaging input, and image** refers to considering the lesions features individually. image\_avg refers to using the average of the lesions' feature vectors as input, while image\_indiv models consider the lesions' features individually.
}
\begin{center}
\begin{tabular}{|l|c|c|c|r|}
 \hline
 Model & Clinical data & Image data & AUC & p-value \\
 \hline
 MLP & x & - & $0.66\pm0.04$ & $0.002$ \\
 MLP & x & x (average) &$0.61\pm0.04$ & $<0.001$ \\
 MLP & - & x (average) &$0.47\pm0.04$ & $<0.001$ \\
 MIL & - & x (per lesion) &$0.56\pm0.06$ & $<0.001$ \\
 GraphConv & - & x (per lesion) & $0.58\pm0.06$ & $<0.001$ \\
 Cross-attention & x & x (per lesion) & $\mathbf{0.72\pm0.03}$ & --- \\
\hline
\end{tabular}
\end{center}
\end{table}
\label{table_results}
\vspace{-1cm}
% \begin{table}
% \begin{tabular}{p{1.2cm}|p{1.3cm} p{1.5cm} p{1.2cm} p{1.2cm} p{1.8cm} p{1.4cm} c} 
%  \hline
%  Model & \centering MLP & \centering MLP & \centering MLP & \centering MIL & \centering GraphConv & \centering GATv2 &  Cross-attention \\ 
%  \hline
%  Input  & \centering clinical & \centering clinical+ & \centering image* & \centering image** & \centering image** & \centering image** &  clinical+ \\ 
% \centering  data &  & \centering image* &  &  &  &  &  image** \\ 
%  \hline
%  %AUC & \underline{0.66+/-0.04} &  \underline{0.66+/-0.04} & 0.47+/-0.04 & 0.56+/-0.06 & 0.58+/-0.06 & 0.46+/-0.07 & \textbf{0.72+/-0.03} \\ 
%  AUC & \centering 0.66 &  \centering 0.61 & \centering 0.47 & \centering 0.56 & \centering 0.58 & \centering 0.46 &  \textbf{0.72} \\ 
%   & \centering $\pm$ 0.04 &  \centering $\pm$0.04 & \centering $\pm$0.04 & \centering $\pm$0.06 & \centering $\pm$0.06 & \centering $\pm$0.07 &  \textbf{$\pm$0.03} \\ 
%  p-value & \centering 0.002 & \centering <0.001 & \centering <0.001 & \centering <0.001 & \centering <0.001 & \centering <0.001 & - \\ 
%  \hline
% \end{tabular}
% \end{table} %question du GAT

%Mettre tableau avec configs ds supplementary material

\subsubsection{Qualitative results}
We also studied the learned attention weights in the cross-attention modules (c.f. figures in %examples in 
Supp. material) in order to 
%ensure that the proposed framework focuses on the right information 
better understand where the model focuses when learning to predict the patients' 2-year PFS. Firstly, we observe that 
%on both cross-attention modules, %all 
%the nodes focus on the same information with the same (or really close) attention weights  
the cross-attention weights across patients can behave differently, with either overall constant weights across rows (lesions) and columns (clinical variables), or approximately constant rows, or variations across rows and columns. 
%(c.f.  examples in the 
%Supp material
%Supp. material~4).
However, the two cross-attention modules for a patient tend to be similar.
%This could mean that the GATv2 creates an homogeneous representation on the nodes of the graph. 
Secondly, the contribution of the different clinical features is mostly %quite often constant 
%(or still very close)
equilibrated: each clinical feature is given approximately the same amount of attention, which is expected since, as we mentioned before, we rely on known biomarkers. %This may be pertinent for patient whose imaging data brings more information for the prediction than the clinical data. 
For some patients, few clinical features stand out. For example, for one patient (Fig.~2 in Supp. material), the model puts a strong attention on his LDH value, which is quite low, and on his aaIPI (age-adjusted International Prognostic Index, which is equal to 1). The prediction for this patient is negative, i.e., no relapse within two years.
%is that he won't have a relapse before two years: 
%this hypothesis is indeed supported
%The prediction is supported by the two clinical variables
%Two clinical values detain the highest attention,
%values 
%the model puts attention on, %Thus, the attention process seems in this case clinically
This seems coherent with the physician's thinking process when trying to asses the condition of a patient, %making us 
confirming the relevance of the multimodal fusion by the cross-attention module.

\section{Conclusion}
%We proposed a new model that integrates both multiple lesions and multiple modalities (dataset composed of clinical and imaging data (FDG-PET/CT)) to improve the treatment response prediction for a patient suffering from DLBCL. These informations allows it to perform significantly better than all baseline models, by learning relevant links between the multimodal data. As perspectives, we would like to consider a cost function adapted to the specific task of survival analysis, which would precise the treatment response estimation (by computing a regression rather than a binary classification). We also want to test the method from \cite{Lv_Zhou_Peng_Peng_Lin_Wu_Xu_Lu_2023} that define the lesion graph on sub-regions of the lesions rather than on the whole lesions, which could help to mitigate the impact of the intra/inter-operator segmentation variability by automatically dividing (or not) lesions whose delimitation is unclear. Finally, we want to test our model on datasets from other pathologies that include a multiplicity of lesions, such as the multiple myeloma, to prove its generalisation ability.
We address treatment response prediction of DLBCL patients two years after diagnosis. To this end, we propose a new cross-attention graph learning method integrating image information from multiple lesions and clinical tabular data. Experimental validation on a prospective clinical dataset shows that our model can efficiently exploit the complementary information, performing significantly better than all compared baselines. 
As perspectives, we will consider cost functions adapted to survival analysis for a more fine-grained treatment response estimation in time and a better modelling of censored patients. In addition, studying graphs defined on lesions sub-regions rather than whole lesions~\cite{Lv_Zhou_Peng_Peng_Lin_Wu_Xu_Lu_2023} could help mitigate the impact of intra/inter-operator segmentation variability, especially for lesions whose delimitation is unclear. Finally, we plan to investigate the generalisation ability of our model to other pathologies.

% \dm{Commented the acknowledgements as it is double blind}
\subsubsection{Acknowledgements} \small This work has been funded by the AIby4 project (Centrale Nantes-Project ANR-20-THIA-0011), INCa-DGOS-INSERM-ITMO Cancer 18011 (SIRIC ILIAD) with the support from the Pays de la Loire region (GCS IRECAN 220729), the European Regional Development Fund (FEDER), the Pays de la Loire region on the Connect Talent MILCOM programme  and Nantes Métropole (Conv. 2017-10470).

\bibliographystyle{splncs04}
\bibliography{biblio-macros,bibliography2}

\end{document}